\documentclass[11pt,a4paper,twoside]{article}
\usepackage{graphicx}
\usepackage{color}
\usepackage{hyperref}
\begin{document}
\baselineskip=0.20in \vspace{20mm} \baselineskip=0.30in
{\bf \LARGE
\begin{center}
Approximate relativistic bound states of a particle in Yukawa field with Coulomb tensor interaction
\end{center}}
\vspace{3mm}
\begin{center}
{\Large {\bf Sameer M. Ikhdair}}\footnote{\scriptsize E-mail:~ sikhdair@neu.edu.tr;~ sikhdair@gmail.com.} \\
\end{center}
{\small
\begin{center}
{\it Physics Department, Near East University, 922022 Nicosia, Northern Cyprus, Turkey.}\\
and\\
{\it Physics Department, Faculty of Science, An-Najah National University, \\
Nablus, West Bank, Palestine.}
\end{center}}
\vspace{3mm}
\begin{center}
{\Large {\bf Babatunde J. Falaye}}\footnote{\scriptsize E-mail:~ fbjames11@physicist.net} \\
\end{center}
{\small
\begin{center}
{\it Theoretical Physics Section, Department of Physics\\
 University of Ilorin,  P. M. B. 1515, Ilorin, Nigeria.}
\end{center}}
\vspace{10mm}
\noindent
\begin{abstract}
\noindent
We obtain the approximate relativistic bound state of a spin-$1/2$ particle in the field of the Yukawa potential and a Coulomb-like tensor interaction with arbitrary spin-orbit coupling number $\kappa$   under the spin and pseudospin (p-spin) symmetries. The asymptotic iteration method is used to obtain energy eigenvalues and corresponding wave functions in their closed forms. Our numerical results show that the tensor interaction removes degeneracies between the spin and p-spin doublets and creates new degenerate doublets for various strengths of tensor coupling.
\end{abstract}

{\bf Keywords}: Dirac equation; Yukawa potential; Coulomb-like tensor interaction; spin and

p-spin symmetries, asymptotic iteration method

{\bf PACS:} 03.65.Ge, 03.65.Fd, 03.65.Pm, 02.30.Gp
\section{Introduction}
Within the framework of Dirac equation, the p-spin symmetry is used to feature deformed nuclei, superdeformation and to establish an effective shell-model \cite{BJ1, BJ2, BJ3, BJ4}. However, the spin symmetry is relevant for mesons \cite{BJ5}. The spin symmetry occurs when $\Delta(r)=V(r)-S(r)=$constant or the scalar potential $S(r)$ is nearly equal to the vector potential $V(r)$, i.e., $S(r)\approx V(r)$. The p-spin symmetry occurs when $\Sigma(r)=V(r)+S(r)=$constant or $S(r)\approx -V(r)$ \cite{BJ6, BJ7, BJ7A, BJ7B}. 

Furthermore, the p-spin symmetry refers to a quasi-degeneracy of single nucleon doublets with non-relativistic quantum number ($n$, $\ell$, $j=\ell+1/2$) and ($n-1$, $\ell+2$, $j=\ell+3/2$), where $n$, $\ell$ and $j$ are single nucleon radial, orbital and total angular quantum numbers, respectively $\cite{BJ8, BJ9}$. The total angular momentum is $j=\tilde{\ell}+\tilde{s}$, where $\tilde{\ell}=\ell+1$ is pseudo-angular momentum and $\tilde{s}$ is p-spin angular momentum $\cite{BJ1}$. Recently, the tensor potential was introduced into the Dirac equation with the substitution: $\vec{P}\rightarrow \vec{P}-im\omega\beta.\hat{r}U(r)$ and a spin-orbit coupling is added to the Dirac Hamiltonian $\cite{BJ10, BJ11, BJ12, BJ13, BJ14, BJ15}$. 

The Yukawa potential or static screened Coulomb potential (SSCP) $\cite{BJ16, NEW1, NEW2}$ can be defined as
\begin{equation} 
V(r)=-V_0\frac{e^{-ar}}{r},
\label{EA}
\end{equation}
where $V_0=\alpha Z$, $\alpha=(137.037)^{-1}$ is the fine-structure constant, $Z$ is the atomic number and $a$ is the screening parameter. This potential is often used to compute bound-state normalizations and energy levels of neutral atoms $\cite{BJ17, BJ18, BJ19}$. Over the past years, several methods have been used in solving relativistic and nonrelativistic equations with the Yukawa potential such as the shifted large-method $\cite{BJ20}$, perturbative solution of the Riccati equation $\cite{BJ21, BJ22}$, alternative perturbative scheme $\cite{BJ23, BJ24}$, the quasi-linearization method (QLM) $\cite{BJ25}$ and Nikiforov-Uvarov method \cite{BJ26}.

The tensor coupling, a higher order term in a relativistic expansion, increases significantly the spin-orbit coupling. This suggests that the tensor coupling could have a significant contribution to p-spin splittings in nuclei as well. This contribution is expected to be particularly relevant for the levels near the Fermi surface, because the tensor coupling depends on the derivative of a vector potential, which has a peak near the Fermi surface for typical nuclear mean-field vector potentials. It has also been used as a natural way to introduce the harmonic oscillator in a relativistic (Dirac) formalism. In a recent article, it was shown that the harmonic oscillator with scalar and vector potentials can exhibit an exact p-spin symmetry \cite{BJ12, NN1}. When this symmetry is broken $(\sum\neq0)$, the breaking term is quite large, manifesting its nonperturbative behavior. However, if a tensor coupling is introduced, the form of harmonic-oscillator potential can still be maintained with $(\sum=0)$, but the p-spin symmetry is broken perturbatively \cite{NN2}.

The tensor interaction has also been considered to explain how the spin-orbit term can be small for $\Lambda-$nucleus and large in the nucleon-nucleus case \cite{NN3}. It is assumed that in the strange sector (case of $\Lambda$ ) the tensor coupling is large and the spin-orbit term obtained from this interaction can cancel in part the contribution coming from the scalar and vector interactions. This result shows that the tensor interaction can change strongly the spin-orbit term. 

It is therefore the aim of the present work is to apply the asymptotic iteration method \cite{BJ27, BJ28, N4, BJ29} in solving the Dirac equation with the Yukawa potential including a Coulomb-like tensor interaction to obtain the energy eigenvalues and corresponding wave functions in view of spin and p-spin symmetry.

The paper is organized as follows. In Section \ref{MOA2}, the AIM is briefly introduced. In Section \ref{TDE3}, we present the Dirac equation with scalar and vector potentials for  arbitrary spin-orbit coupling number   including tensor interaction in view of spin and p-spin symmetry. In Section \ref{ARS4}, we obtain the energy eigenvalue equations and corresponding wave functions. We discuss our numerical results in Section $\ref{NR5}$. Our conclusion is given in Section $\ref{C6}$.

\section{Method of Analysis}
\label{MOA2}
One of the calculational tools utilized in solving the Schr\"{o}dinger
dinger-like equation including the centrifugal barrier and/or the spin-orbit
coupling term is called as the asymptotic iteration method (AIM). For a
given potential the idea is to convert the Schr\"{o}dinger-like
equation to the homogenous linear second-order differential equation of the
form: 
\begin{equation}
y^{\prime\prime}(x)=\lambda_o(x)y^{\prime}(x)+s_o(x)y(x),  \label{E1}
\end{equation}
where $\lambda_o(x)$ and $s_o(x)$ have sufficiently many continous
derivatives and defined in some interval which are not necessarily bounded.
The differential Eq. (\ref{E1}) has a general solution $\cite{BJ27,BJ8}$ 
\begin{equation}
y(x)=\exp\left(-\int^x\alpha(x^{\prime})dx^{\prime}\right)\left[
C_2+C_1\int^x\exp\left(\int^{x^{\prime}}\left[\lambda_o(x^{\prime\prime})+2
\alpha(x^{\prime\prime})\right]dx^{\prime\prime}\right)dx^{\prime}\right].
\label{E2}
\end{equation}
If $k>0$, for sufficiently large $k$, we obtain the $\alpha(x)$ 
\begin{equation}
\frac{s_k(x)}{\lambda_k(x)}=\frac{s_{k-1}(x)}{\lambda_{k-1}(x)}=\alpha(x) \
,\ \ k=1, 2, 3.....  \label{E3}
\end{equation}
where
\begin{eqnarray} 
\lambda_k(x)&=&\lambda^{\prime}_{k-1}(x)+s_{k-1}(x)+\lambda_o(x)
\lambda_{k-1}(x),\nonumber\\
s_k(x)&=&s^{\prime}_{k-1}(x)+s_o(x)\lambda_{k-1}(x) \ ,\ \ k=1, 2, 3.....
\label{E4}
\end{eqnarray}
with quantization condition 
\begin{equation}
\delta_k(x)= \left| 
\begin{array}{lr}
\lambda_k(x) & s_k(x) \\ 
\lambda_{k-1}(x) & s_{k-1}(x)%
\end{array}
\right|=0\ \ ,\ \ \ k=1, 2, 3....  \label{E5}
\end{equation}
The energy eigenvalues are then obtained from (\ref{E5}), if the problem is
exactly solvable.

\section{Dirac Equation with a Tensor Interaction}
\label{TDE3}
In spherical coordinates, the Dirac equation for fermonic massive spin$-1/2$
particles interacting with arbitrary scalar potential $S(r)$, the
time-component $V(r)$ of a four-vector potential and the tensor potential $%
U(r)$ can be expressed as $\cite{BJ1, BJ7}$ 
\begin{equation}
\left[ \vec{\alpha}.\vec{p}+\beta (M+S(r))-i\beta \vec{\alpha}.\hat{r}U(r)%
\right] \psi (\vec{r})=[E-V(r)]\psi (\vec{r}),  \label{E6}
\end{equation}%
where $E$, $\vec{p}$ and $M$ denote the relativistic energy of the system,
the momentum operator and mass of the particle respectively. $\alpha $ and $%
\beta $ are $4\times 4$ Dirac matrices given by 
\begin{equation}
\bar{\alpha}=\left( 
\begin{array}{lr}
0 & \vec{\sigma} \\ 
\vec{\sigma} & 0%
\end{array}%
\right) \ ,\ \ \ \beta =\left( 
\begin{array}{lr}
I & 0 \\ 
0 & -I%
\end{array}%
\right) ,\ \ \ {\sigma _{1}}=\left( 
\begin{array}{lr}
0 & 1 \\ 
1 & 0%
\end{array}%
\right) ,\ \ \ {\sigma _{2}}=\left( 
\begin{array}{lr}
0 & -1 \\ 
i & 0%
\end{array}%
\right) ,\ \ \ {\sigma _{3}}=\left( 
\begin{array}{lr}
1 & 0 \\ 
0 & -1%
\end{array}%
\right) ,  \label{E7}
\end{equation}%
where $I$ is the $2\times 2$ unitary matrix and $\vec{\sigma}$ are the
three-vector pauli spin matrices. The eigenvalues of the spin-orbit coupling
operator are $\kappa =\left( j+\frac{1}{2}\right) >0$ and $\kappa =-\left( j+%
\frac{1}{2}\right) <0$ for unaligned spin $j=\ell -\frac{1}{2}$ and the
aligned spin $j=\ell +\frac{1}{2}$ respectively. The set $%
(H^{2},K,J^{2},J_{Z})$ can be taken as the complete set of conservative
quantities with $\vec{J}$ being the total angular momentum operator and $K=(%
\vec{\sigma}.\vec{L}+1)$ is the spin-orbit where $\vec{L}$ is the orbital
angular momentum of the spherical nucleons that commutes with the Dirac
Hamiltonian. Thus, the spinor wave functions can be classified according to
their angular momentum $j$, the spin-orbit quantum number $\kappa $ and the
radial quantum number $n$. Hence, they can be written as follows: 
\begin{equation}
\psi _{n\kappa }(\vec{r})=\left( 
\begin{array}{lr}
f_{n\kappa }(\vec{r}) &  \\ 
g_{n\kappa }(\vec{r}) & 
\end{array}%
\right) =\frac{1}{r}\left( 
\begin{array}{lr}
F_{n\kappa }(r)Y_{jm}^{\ell }(\theta,\phi ) &  \\ 
iG_{n\kappa }(r)Y_{jm}^{\tilde{\ell}}(\theta,\phi ), & 
\end{array}%
\right) ,  \label{E8}
\end{equation}%
where $F_{n\kappa }(r)$ and $G_{n\kappa }(r)$ are the radial
wave functions of the upper- and lower-spinor components respectively and $%
Y_{jm}^{\ell }(\theta,\phi )$ and $Y_{jm}^{\tilde{\ell}}(\theta,\phi )$ are the
spherical harmonic functions coupled to the total angular momentum $j$ and
it's projection $m$ on the $z$ axis. Substitution of equation (\ref{E6})
into equation (\ref{E1}) yields the following coupled differential
equations: 
\begin{eqnarray}
\left( \frac{d}{dr}+\frac{\kappa }{r}-U(r)\right) F_{n\kappa }(r)
&=&(M-E_{n\kappa }-\Delta (r))G_{n\kappa }(r)  \nonumber \\
\left( \frac{d}{dr}-\frac{\kappa }{r}+U(r)\right) G_{n\kappa }(r)
&=&(M-E_{n\kappa }-\Sigma (r))F_{n\kappa }(r)  \label{E9}
\end{eqnarray}%
where $\Delta (r)=V(r)-S(r)$ and $\Sigma (r)=V(r)+S(r)$ are the difference
and sum potentials respectively. After eliminating $F_{n\kappa }(r)$ and $%
G_{n\kappa }(r)$ in equations (\ref{E9}), we obtain the following two Schr$%
\ddot{o}$diger-like differential equations for the upper and lower spinor
components: 
\begin{eqnarray}
&&\left[ \frac{d^{2}}{dr^{2}}-\frac{\kappa (\kappa +1)}{r^{2}}+\frac{2\kappa 
}{r}U(r)-U^{2}(r)-\frac{dU(r)}{dr}+\frac{\frac{d\Delta (r)}{dr}}{%
M+E_{n\kappa }-\Delta (r)}\left( \frac{d}{dr}+\frac{\kappa }{r}-U(r)\right) %
\right] F_{n\kappa }(r)  \nonumber \\
&=&\left[ \left( M+E_{n\kappa }-\Delta (r)\right) \left( M-E_{n\kappa
}+\Sigma (r)\right) \right] F_{n\kappa }(r),  \label{E10}
\end{eqnarray}%
\begin{eqnarray}
&&\left[ \frac{d^{2}}{dr^{2}}-\frac{\kappa (\kappa -1)}{r^{2}}+\frac{2\kappa 
}{r}U(r)-U^{2}(r)+\frac{dU(r)}{dr}+\frac{\frac{d\Sigma (r)}{dr}}{%
M-E_{n\kappa }+\Sigma (r)}\left( \frac{d}{dr}-\frac{\kappa }{r}+U(r)\right) %
\right] G_{n\kappa }(r)  \nonumber \\
&=&\left[ \left( M+E_{n\kappa }-\Delta (r)\right) \left( M-E_{n\kappa }+\Sigma
(r)\right) \right] G_{n\kappa }(r),  \label{E11}
\end{eqnarray}%
where $\kappa (\kappa -1)=\bar{\ell}(\bar{\ell}+1)$ and $\kappa (\kappa +1)=%
\ell(\ell+1)$

\subsection{P-spin Symmetry Limit}
\label{PSL}
The pseudospin symmetry occurs when $\frac{d[V(r)+S(r)]}{dr}=%
\frac{d\Sigma (r)}{dr}=0$ or $\Sigma (r)=C_{ps}=$constant. Here we are
taking $\Delta (r)$ as the Yukawa potential and the tensor as the
Coulomb-like potential, i.e. 
\begin{equation}
\Delta (r)=-V_{o}\frac{e^{-ar}}{r}\ \ and\ \ U(r)=-\frac{A}{r},\ \ \ \ r\geq
R_{c},  \label{E12}
\end{equation}%
with 
\begin{equation}
A=\frac{Z_{a}Z_{b}e^{2}}{4\pi \epsilon _{o}},  \label{E13}
\end{equation}%
where $R_{c}$ is the Coulomb radius, $Z_{a}$ and $Z_{b}$ respectively,
denote the charges of the projectile $a$ and the target nuclei $b$ $\cite%
{BJ1, BJ7}$. Under this symmetry, equation (\ref{E11}) can easily be
transformed to 
\begin{equation}
\left[ \frac{d^{2}}{dr^{2}}-\frac{\tilde{\delta}}{r^{2}}+\frac{\tilde{\gamma}%
e^{-ar}}{r}-\tilde{\beta}^{2}\right] G_{n\kappa }(r)=0,  \label{E14}
\end{equation}%
where $\kappa =-\tilde{\ell}$ and $\kappa =\tilde{\ell}+1$ for $\kappa <0$
and $\kappa >0$ respectively and 
\begin{eqnarray}
\tilde{\gamma} &=&(E_{n\kappa }-M-C_{ps})V_{0},  \nonumber \\
\tilde{\beta} &=&\sqrt{(M+E_{n\kappa })(M-E_{n\kappa }+C_{ps})}, \\
\tilde{\delta} &=&(\kappa +A)(\kappa +A-1),  \nonumber  \label{E15}
\end{eqnarray}%
have been introduced for simplicity.

\subsection{Spin Symmetry Limit}

\label{SSL} In the spin symmetry limit, $\frac{d\Delta (r)}{dr}=0$ or $%
\Delta (r)=C_{s}=$ constant $\cite{BJ1, BJ7}$. Similarly to section \ref{PSL}%
, we consider 
\begin{equation}
\Sigma (r)=-V_{o}\frac{e^{-ar}}{r}\ \ and\ \ U(r)=-\frac{A}{r},\ \ \ \ r\geq
R_{c}.  \label{E16}
\end{equation}%
With equations (\ref{E16}), equation (\ref{E10}) can be transformed to 
\begin{equation}
\left[ \frac{d^{2}}{dr^{2}}-\frac{{\delta }}{r^{2}}+\frac{{\gamma }e^{-ar}}{r%
}-{\beta }^{2}\right] F_{n\kappa }(r)=0,  \label{E17}
\end{equation}%
where $\kappa =\ell $ and $\kappa =-\ell -1$ for $\kappa <0$ and $\kappa >0$%
, respectively. We have also introduced the following parameters 
\begin{eqnarray}
\gamma  &=&(M+E_{n\kappa }-C_{s})V_{0},  \nonumber \\
\beta  &=&\sqrt{(M-E_{n\kappa })(M+E_{n\kappa }-C_{s})}, \\
\delta  &=&(\kappa +A)(\kappa +A+1),  \nonumber  \label{E18}
\end{eqnarray}%
for simplicity.

\section{Approximate Relativistic Bound States}
\label{ARS4}
In this section, within the framework of the AIM, we shall solve
the Dirac equation with the Yukawa potential in the presence of the tensor
potential.

\subsection{P-Spin Symmetric Solution}
\label{PSC}
 To obtain analytical approximate solution for the Yukawa
potential, an approximation has to be made for the centrifugal term $1/r^{2},
$ which is similar to the one taken by other authors $[37-43]$ 
\begin{equation}
\frac{1}{r^{2}}\approx \frac{4a^{2}e^{-2ar}}{\left( 1-e^{-2ar}\right) ^{2}},
\label{E19}
\end{equation}%
which is valid for $ar<<1$. With this approximation, equation (\ref{E14})
can be written as 
\begin{equation}
\frac{d^{2}G_{n\kappa }(r)}{dr^{2}}+\left[ \frac{2a\tilde{\gamma}e^{-2ar}}{%
\left( 1-e^{-2ar}\right) }-\frac{4a^{2}\tilde{\delta}e^{-2ar}}{\left(
1-e^{-2ar}\right) ^{2}}-\tilde{\beta}^{2}\right] G_{n\kappa }(r)=0.
\label{E20}
\end{equation}%
In order to obtain the solution of equation (\ref{E20}), we introduce a
transformation of the form $z=\left( e^{-2ar}-1\right) ^{-1}$, as a result,
equation (\ref{E20}) can be rewritten as 
\begin{equation}
z(z+1)\frac{d^{2}G_{n\kappa }(z)}{dz^{2}}+(1+2z)\frac{dG_{n\kappa }(z)}{dz}-%
\frac{\left( \tilde{\delta}z^{2}+\left( \tilde{\delta}+\frac{\tilde{\gamma}}{%
2a}\right) z+\left( \frac{\tilde{\gamma}}{2a}+\frac{\tilde{\beta}^{2}}{4a^{2}%
}\right) \right) }{z(1+z)}G_{n\kappa }(r).  \label{E21}
\end{equation}%
According to the Frebenius theorem, the singularity points of the above
differential equation play an essential role in the form of the wave
functions. The singular points of the above equation (\ref{E21}) are at $z=0$
and $z=-1$. As a result, we take the wave functions of the form 
\begin{equation}
G_{n\kappa }(z)=z^{\tilde{\rho}}(1+z)^{-\frac{\tilde{\beta}}{2a}}g_{n\kappa
}(z),  \label{E22}
\end{equation}%
where 
\begin{equation}
\tilde{\rho}=\sqrt{\frac{\tilde{\beta}^{2}}{4a^{2}}+\frac{\tilde{\gamma}}{2a}%
}.  \label{E23}
\end{equation}%
Substituting equations (\ref{E22}) and (\ref{E23}) into equation (\ref{E21})
allows us to find the following second-order equation 
\begin{equation}
\frac{d^{2}g_{n\kappa }(z)}{dz^{2}}-\left[ \frac{2z\left( \frac{\tilde{\beta}%
}{2a}-\tilde{\rho}-1\right) -(2\tilde{\rho}+1)}{z(1+z)}\right] \frac{%
dg_{n\kappa }}{dz}-\left[ \frac{\left( \frac{\tilde{\beta}}{2a}-\tilde{\rho}%
\right) -\left( \frac{\tilde{\beta}}{2a}-\tilde{\rho}\right) ^{2}+\tilde{%
\delta}}{z(z+1)}\right] g_{n\kappa }(z),  \label{E24}
\end{equation}%
which is suitable to an AIM solutions. In order to use the AIM procedure, we
compare equation (\ref{E24}) with equation (\ref{E1}) and obtain $\lambda
_{0}(z)$ and $s_{0}(z)$ equations as 
\begin{eqnarray}
\lambda _{0}(z) &=&\frac{2z\left( \frac{\tilde{\beta}}{2a}-\tilde{\rho}%
-1\right) -(2\tilde{\rho}+1)}{z(1+z)},  \nonumber \\
s_{0}(z) &=&\frac{\left( \frac{\tilde{\beta}}{2a}-\tilde{\rho}\right)
-\left( \frac{\tilde{\beta}}{2a}-\tilde{\rho}\right) ^{2}+\tilde{\delta}}{%
z(z+1)}.  \label{E25}
\end{eqnarray}%
By using the termination condition of the AIM given in equation (\ref{E5}),
we obtain 
\begin{eqnarray}
\delta _{0}(z) &=&\left\vert 
\begin{array}{lr}
\lambda _{1}(z) & s_{1}(z) \\ 
\lambda _{0}(z) & s_{o}(z)%
\end{array}%
\right\vert =0\ \ \ \ \Rightarrow \ \ \ \ \tilde{\rho}_{0}-\frac{\tilde{%
\beta _{0}}}{2a}=-\frac{1}{2}-\frac{1}{2}\sqrt{1+4\tilde{\delta}}  \nonumber
\\
\delta _{1}(z) &=&\left\vert 
\begin{array}{lr}
\lambda _{2}(z) & s_{2}(z) \\ 
\lambda _{1}(z) & s_{1}(z)%
\end{array}%
\right\vert =0\ \ \ \ \Rightarrow \ \ \ \ \tilde{\rho}_{1}-\frac{\tilde{%
\beta _{1}}}{2a}=-\frac{3}{2}-\frac{1}{2}\sqrt{1+4\tilde{\delta}} \\
\delta _{2}(z) &=&\left\vert 
\begin{array}{lr}
\lambda _{3}(z) & s_{3}(z) \\ 
\lambda _{2}(z) & s_{2}(z)%
\end{array}%
\right\vert =0\ \ \ \ \Rightarrow \ \ \ \ \tilde{\rho}_{2}-\frac{\tilde{%
\beta _{2}}}{2a}=-\frac{5}{2}-\frac{1}{2}\sqrt{1+4\tilde{\delta}}  \nonumber
\\
&&\ldots etc.  \nonumber  \label{E26}
\end{eqnarray}%
The above expressions can be generalized as 
\begin{equation}
\tilde{\rho}_{n}-\frac{\tilde{\beta _{n}}}{2a}=-\frac{2n+1}{2}-\frac{1}{2}%
\sqrt{1+4\tilde{\delta}}.  \label{E27}
\end{equation}%
If one insert the values of $\tilde{\rho}$, $\tilde{\beta}$ and $\tilde{%
\delta}$ into equation (\ref{E27}), the energy spectrum equation can be
obtained as 
\begin{equation}
\sqrt{\left( M+E_{n\kappa }\right) \left( M-E_{n\kappa }+C_{ps}\right) }-2a(\kappa +A+n)=%
\sqrt{\left( M-E_{n\kappa }+C_{ps}\right) \left( M+E_{n\kappa
}-2aV_{o}\right) }.  \label{E28}
\end{equation}%
On squaring both sides of equation (\ref{E28}), we obtain a more explicit
expression for the energy spectrum as 
\begin{equation}
\frac{\left[ 2a(n+\kappa +A)^{2}+V_{o}(M-E_{n\kappa }+C_{ps})\right] ^{2}}{4(M+E_{n\kappa
})(M-E_{n\kappa }+C_{ps})}=(n+\kappa +A)^{2}.  \label{E28b}
\end{equation}%
For special case when $A=0$, our result is exacly identical with the one
obtained by Ikhdair {\cite{BJ33}} and by Aydo$\check{g}$du and Sever {\cite%
{BJ34}}. Now we shall obtain the eigenfunction using the AIM. Generally
speaking, the differential equation we wish to solve should be transformed
to the form $\cite{BJ28}$: 
\begin{equation}
y^{\prime \prime }(x)=2\left( \frac{\Lambda x^{N+1}}{1-bx^{N+2}}-\frac{m+1}{x%
}\right) y^{\prime }(x)-\frac{Wx^{N}}{1-bx^{N+2}},  \label{E29}
\end{equation}%
where $a$, $b$ and $m$ are constants. The general solution of equation (\ref%
{E29}) is found as {\cite{BJ2}} 
\begin{equation}
y_{n}(x)=(-1)^{n}C_{2}(N+2)^{n}(\sigma )_{_{n}}{_{2}F_{1}(-n,t+n;\sigma
;bx^{N+2})},  \label{E30}
\end{equation}%
where the following notations have been used 
\begin{equation}
(\sigma )_{_{n}}=\frac{\Gamma {(\sigma +n)}}{\Gamma {(\sigma )}}\ \ ,\ \
\sigma =\frac{2m+N+3}{N+2}\ \ and\ \ \ t=\frac{(2m+1)b+2\Lambda }{(N+2)b}.
\label{E31}
\end{equation}%
Now, comparing equations (\ref{E30}) and (\ref{E21}), we have $\Lambda =%
\frac{\tilde{\beta}}{2a}-\frac{1}{2}$, $b=-1$, $N=-1$, $m=\tilde{\rho}-\frac{%
1}{2}$, $\sigma =2\tilde{\rho}+1$, $t=2\left( \tilde{\rho}-\frac{\tilde{\beta%
}}{2a}\right) $ and then the solution of equation (\ref{E24}) can easily
found as 
\begin{equation}
g_{n\kappa }(z)=(-1)^{n}\frac{\Gamma (2\tilde{\rho}+1)}{\Gamma (2\tilde{\rho}%
)}\ {_{2}F_{1}}\left( -n,2\left( \tilde{\rho}-\frac{\tilde{\beta}}{2a}%
\right) +n+1;2\tilde{\rho}+1;-z\right) ,  \label{E32}
\end{equation}%
where $\Gamma $ and $_{2}F_{1}$ are the Gamma function and hypergeometric
function, respectively. By using equations (\ref{E22}) and (\ref{E32}) we
can write the corresponding lower spinor component $G_{n\kappa }(z)$ as 
\begin{equation}
G_{n\kappa }(z)=N_{n\kappa}z^{\sqrt{\frac{\tilde{\beta}^{2}}{4a^{2}}+%
\frac{\tilde{\gamma}}{2a}}}(1+z)^{-\frac{\tilde{\beta}}{2a}}\ {_{2}F_{1}}%
\left( -n,2\left( \sqrt{\frac{\tilde{\beta}^{2}}{4a^{2}}+\frac{\tilde{\gamma}%
}{2a}}-\frac{\tilde{\beta}}{2a}\right) +n+1;2\sqrt{\frac{\tilde{\beta}^{2}}{%
4a^{2}}+\frac{\tilde{\gamma}}{2a}}+1;-z\right) ,  \label{E33}
\end{equation}%
where $N_{n\kappa}$ is the normalization constant.

\subsection{Spin Symmetric Solution}
\label{SSC}
 Following the previous section \ref{PSC}, approximation (\ref%
{E19}) is used instead of the centrifugal term $1/r^{2}$ and we rewrite
equation (\ref{E17}) as 
\begin{equation}
\frac{d^{2}F_{n\kappa }(r)}{dr^{2}}+\left[ \frac{2a{\gamma }e^{-2ar}}{\left(
1-e^{-2ar}\right) }-\frac{4a^{2}{\delta }e^{-2ar}}{\left( 1-e^{-2ar}\right)
^{2}}-{\beta }^{2}\right] F_{n\kappa }(r)=0.  \label{E34}
\end{equation}%
We have decided to use the same variable so as to avoid repetition of
algebra. It is clear that equation (\ref{E34}) is similar to equation (\ref%
{E20}); therefore substituting for $\gamma $, $\delta $ and $\beta $ in
equation (\ref{E27}), the relativistic energy spectrum turns out as 
\begin{equation}
\frac{\left[ 2a(n+\kappa +A+1)^{2}-V_{0}(M+E_{n\kappa }-C_{s})\right] ^{2}}{4(M-E_{n\kappa
})(M+E_{n\kappa }-C_{s})}=(n+\kappa +A+1)^{2},  \label{E35}
\end{equation}%
and the associated upper spinor component $F_{n\kappa }(z)$ as 
\begin{equation}
F_{n,\kappa }(z)=C_{n,\kappa}z^{\sqrt{\frac{{\beta }^{2}}{4a^{2}}+\frac{{%
\gamma }}{2a}}}(1+z)^{-\frac{{\beta }}{2a}}\ {_{2}F_{1}}\left( -n,2\left( 
\sqrt{\frac{{\beta }^{2}}{4a^{2}}+\frac{{\gamma }}{2a}}-\frac{{\beta }}{2a}%
\right) +n+1;2\sqrt{\frac{{\beta }^{2}}{4a^{2}}+\frac{{\gamma }}{2a}}%
+1;-z\right) ,  \label{E36}
\end{equation}%
where $C_{n,\kappa}$ is the normalization constant.

Unlike the non relativistic case,  the normalization condition for the Dirac spinor combines the two individual normalization constants $N_{n,\kappa}$ and $C_{n,\kappa}$ in one single integral. The radial wave functions are normalized according to the formula $\int_{0}^{\infty} \psi ^{\dagger}\psi r^2 dr=1$ which explicitly implies for $f_{n,\kappa}(r)$ and $g_{n,\kappa}(r)$ in equation (\ref{E8}) that {\cite{BJ36}}
\begin{equation}
\int_{0}^{\infty} \left( f_{n,\kappa}^{2}(r)+g_{n,\kappa}^{2}(r)\right) r^{2} dr=\int_{0}^{\infty} \left( F_{n,\kappa}^{2}(r)+G_{n,\kappa}^{2}(r)\right) dr=1.  \label{E37}
\end{equation}%
where the upper and lower spinor components of the total radial wave functions can be expressed in terms of the confluent hypergeometric functions as \cite{BJ33}
\begin{equation}
F_{n\kappa}(s)=N\frac{\Gamma(n+2\beta+1)}{\Gamma(2\beta+1)n!}s^\beta(1-s)^{\kappa+A+1}\ _2F_1\left(-n, 2(\beta+\kappa+A+1)+n;2\beta+1; s\right),
\label{E38}
\end{equation}
and 
\begin{equation}
G_{n\kappa}(s)=N\frac{\Gamma(n+2\gamma+1)}{\Gamma(2\gamma+1)n!}s^\gamma(1-s)^{\kappa+A}\ _2F_1\left(-n, 2(\gamma+\kappa+A+1)+n; 2\gamma+1; s\right),
\label{E39}
\end{equation}
with
\begin{equation}
\beta=\sqrt{(M-E_{n\kappa})(M+E_{n\kappa}-C_s)},\ \ \ \  \gamma=\sqrt{(M+E_{n\kappa})(M-E_{n\kappa}+C_{ps})}, \ \ \ s=e^{-2a r}
\label{E40}.
\end{equation}
Hence, equation (\ref{E37}) can be re-written in terms of variable $s$ as
\begin{equation}
\int^1_0\frac{ds}{s}\left[F_{n,\kappa}^2(s)+G_{n\kappa}^2(s)\right]=2a,
\label{E41}
\end{equation}
where $s\rightarrow 1$ when $r\rightarrow 0$ and $s\rightarrow 0$ when $r\rightarrow \infty$. The method to compute $N$ is given in Ref. \cite{Saad}. Thus we have
\begin{eqnarray}
N^2\left[\left(\frac{\Gamma(n+2\beta+1)}{\Gamma(2\beta+1)n!}\right)^2\int^1_0s^{2\beta-1}(1-s)^{2(\kappa+A+1)}\left[ _2F_1 \left(-n, 2(\beta+\kappa+A+1)+n;2\beta+1; s\right)\right]^2ds\right]\nonumber\\
+N^2\left[\left(\frac{\Gamma(n+2\gamma+1)}{\Gamma(2\gamma+1)n!}\right)^2\int^1_0s^{2\gamma-1}(1-s)^{2(\kappa+A)}\left[_2F_1 \left(-n, 2(\gamma+\kappa+A)+n;2\gamma+1; s\right)\right]^2ds\right]=2a,
\label{E42}
\end{eqnarray}
with the confluent hypergeometric function which is defined by
\begin{equation}
_pF_q\left(a_1,......a_p; b_1,.....b_q; s\right)=\sum_{i=0}^\infty\frac{(a_1)_i...(a_p)_is^i}{(b_1)_i...(b_q)_ii!},
\label{E43}
\end{equation}
where $(a_1)_i$ and $(b_1)_i$ are Pochhammer symbols. We can obtain the normalization constant as \cite{ArWe}:
\begin{eqnarray}
&&N^2\left\{\left(\frac{\Gamma(n+2\beta+1)}{\Gamma(2\beta+1)n!}\right)^2\sum_{i=0}^\infty\frac{(-n)_i(2\beta+2\kappa+2A+2+n)_i(2\beta)_i}{(2\beta+1)_i(2\beta+2\kappa+2A+3)_ii!}B(2\beta,2\kappa+2A+3)\right.\nonumber\\
&&\times\ _3F_2(-n, 2(\beta+\kappa+A+1)+n, 2\beta+i; 2\beta+1, 2(\beta+\kappa+A+1)+1+i; 1)\\
&&+\left.\left(\frac{\Gamma(n+2\gamma+1)}{\Gamma(2\gamma+1)n!}\right)^2\sum_{i=0}^\infty\frac{(-n)_i(2\gamma+2\kappa+2A+n)_i(2\gamma)_i}{(2\gamma+1)_i(2\gamma+2\kappa+2A)_ii!}B(2\gamma,2\kappa+2A+1)\right.\nonumber\\
&&\left.\times\ _3F_2(-n, 2(\gamma+\kappa+A)+n, 2\gamma+i; 2\gamma+1, 2(\gamma+\kappa+A)+1+i; 1)\right\}=2a,\nonumber
\label{E44}
\end{eqnarray}
where 
\begin{eqnarray}
&&B(x,y)=\int^1_0 t^{x-1}(1-t)^{y-1}dt=\frac{\Gamma(x)\Gamma(y)}{\Gamma(x+y)}, \ \ \ \mbox{\textbf{Re}}(x), \mbox{\textbf{Re}}(y)>0,\nonumber\\
&&B\left(\frac{1}{2}, \frac{1}{2}\right)=\pi, B(x,y)=B(y,x).
\label{E45}
\end{eqnarray}
Incomplete beta function is given as
\begin{equation}
B(x; a, b)=\int_0^xt^{a-1}(1-t)^{b-1}dt,
\label{E46}
\end{equation}
and the regularized beta function as
\begin{equation}
I_x(a, b)=\frac{B(x; a, b)}{B(a, b)}
\label{E47}
\end{equation}
In the special case when $A=0$, our result is identical to the one obtained by Ikhdair {\cite{BJ33}} and also by Setare and Haidari $\cite{BJ35}$ by means of the Nikiforov-Uvarov method.
 \section{Numerical Results}
\label{NR5} 
By taking $V_0=1$ and $C_{ps}=-5.0fm^{-1}$, we found that the particle is strongly attracted to the nucleus. In the absence of tensor interaction, i.e., $A=0$, we noticed that the set of p-spin symmetry doublets: 
$\left(1s_{1/2},2p_{3/2}\right)$, $\left(1p_{3/2}, 2s_{1/2}, 2d_{5/2}\right)$, $\left(1d_{5/2}, 2f_{7/2}\right)$, $\left(0d_{3/2},1f_{7/2}\right)$, $\left(0f_{5/2}, 1d_{3/2}\right)$, $\left(0g_{7/2}, 1f_{5/2}\right)$ and $\left(0h_{9/2}, 1g_{7/2}\right)$ have same energies as $-4.500000000$, $-4.497559159$, $-4.490219111$, $-4.477926642$, $-4.460590892$, $-4.438679893$ and $-4.410215299fm^{-1}$, respectively.

It is also noticed that the presence of tensor interaction, say $A=0.5$, removes the degeneracy between the states in the above doublets and creates a new set of p-spin symmetric doublets $\left(1s_{1/2},1p_{3/2},2p_{3/2},2d_{5/2}\right)$, $\left(1d_{5/2}, 2f_{7/2}, 2s_{1/2}\right)$, $\left(1f_{7/2}\right)$, $\left(0d_{3/2}\right)$,$\left(0f_{5/2},1d_{3/2}\right)$, $\left(0g_{7/2}, 1f_{5/2}\right)$ and $\left(0h_{9/2}, 1g_{7/2}\right)$ having identical energies as $-4.499390062$, $-4.494504009$, $-4.484696693$, $-4.469896410$, $-4.449992282$, $-4.424829942$ and $-4.394205177fm^{-1}$, respectively. 

When the strength of tensor interaction increasing, say $A=1.0$, then the following p-spin doublets have same energies as the counter ones when $A=0$ as $\left(1p_{3/2},2d_{5/2}\right)\leftrightarrow\left(1s_{1/2},2p_{3/2}\right)$, $\left(1s_{1/2}, 1d_{5/2}, 2p_{3/2},2f_{7/2}\right)\leftrightarrow
\left(1p_{3/2}, 2s_{1/2}, 2d_{5/2}\right)$,  $\left(1f_{7/2}, 2s_{1/2}\right)\leftrightarrow\left(1d_{5/2}, 2f_{7/2}\right)$, $\left(0d_{3/2}\right)\leftrightarrow\left(0f_{5/2}, 1d_{3/2}\right)$, $\left(0f_{5/2}, 1d_{3/2}\right)\leftrightarrow\left(0g_{7/2}, 1f_{5/2}\right)$ and $\left(0g_{7/2}, 1f_{5/2}\right)\leftrightarrow\left(0h_{9/2}, 1g_{7/2}\right)$. 

Furthermore, we considered the case where $C_{ps}=0$ and the same set of p-spin symmetry doublets: $\left(1s_{1/2},2p_{3/2}\right)$, $\left(1p_{3/2},2s_{1/2},2d_{5/2}\right)$, $\left(1d_{5/2},2f_{7/2}\right)$, $\left(0d_{3/2},1f_{7/2}\right)$, $\left(0f_{5/2},1d_{3/2}\right)$, $\left(0g_{7/2},1f_{5/2}\right)$ and $\left(0h_{9/2},1g_{7/2}\right)$ have same energies as $-4.500000000$, $-0.208711915$, $-0.300000000$, $-0.260536466$, $-0.140000000$, $-2.823651852$ and $-5.111034483fm^{-1}$, respectively whereas the doublet set $\left(1s_{1/2},2p_{3/2}\right)$ has no bound negative energy.

When the tensor strength $A=0.5$, the degeneracy is changed as $0d_{3/2}$, $\left(1d_{5/2},2s_{1/2},2f_{7/2}\right)$, $1f_{7/2}$, $\left(0f_{5/2},1d_{3/2}\right)$, $\left(0g_{7/2},1f_{5/2}\right)$ and $\left(0h_{9/2},1g_{7/2}\right)$ have the bound energies $-0.213820459$, $-0.284176740$, $-0.2892240312$, $-0.586585366$, $-4.074227224$ and $-6.043369573$, respectively, whereas the set $\left(1s_{1/2},1p_{3/2},2d_{5/2},2p_{3/2}\right)$ has no negative bound energy. The particle becomes strongly bounded as $n$ and $\kappa$ increasing. 

By increasing the tensor strength as $A=1$,  the following p-spin doublets have same energies as in case $A=0$: $0d_{3/2}\rightarrow-0.140000000fm^{-1}$, $\left(1s_{1/2},1d_{5/2},2p_{3/2},2f_{7/2}\right)\leftrightarrow\left(1p_{3/2},2s_{1/2},2d_{5/2}\right)$, $\left(1f_{7/2},2s_{1/2}\right)\leftrightarrow\left(1d_{5/2},2f_{7/2}\right)$, $\left(0f_{5/2},1d_{3/2}\right)\leftrightarrow\left(0g_{7/2},1f_{5/2}\right)$ and $\left(0g_{7/2},1f_{5/2}\right)\leftrightarrow\left(0h_{9/2},1g_{7/2}\right)$. The p-spin doublet set $\left(1p_{3/2},2d_{5/2}\right)$ has no negative bound state energy.

Let us also discuss the energy states in the presence of spin symmetry for $C_s=5.0fm^{-1}$ and $C_s=0fm^{-1}$, respectively. In the absence of tensor interaction and when  $C_s=5.0fm^{-1}$,  the degeneracy in the sets $\left(0s_{1/2},1p_{3/2},2d_{5/2},3f_{7/2}\right)$, $\left(1s_{1/2},0p_{3/2},2p_{3/2},1d_{5/2},3d_{5/2},2f_{7/2}\right)$,
$\left(2f_{5/2},3d_{3/2}\right)$
$\left(2s_{1/2},2p_{3/2},0d_{5/2},3d_{5/2},1f_{7/2},0p_{1/2}\right)$,  $\left(3s_{1/2},1p_{1/2},0f_{7/2},0d_{3/2}\right)$, $\left(2p_{1/2},0f_{5/2},1d_{3/2}\right)$, and \\$\left(3p_{1/2},1f_{5/2},2d_{3/2}\right)$, are shown.
 
Furthermore, in the presence of tensor interaction $A=1$, the following doublet sets have identical energies as their counter ones in the absence of tensor:\\ $\left(0p_{3/2},1d_{5/2},2f_{7/2}\right)\leftrightarrow\left(0s_{1/2},1p_{3/2},2d_{5/2},3f_{7/2}\right)$,\\ $\left(0s_{1/2},1p_{3/2},0d_{5/2},2d_{5/2},1f_{7/2},3f_{7/2}\right)\leftrightarrow\left(1s_{1/2},0p_{3/2},2p_{3/2},1d_{5/2},3d_{5/2},2f_{7/2}\right)$,\\ $\left(1s_{1/2},2p_{3/2},3d_{5/2},0f_{7/2}\right)\leftrightarrow\left(2s_{1/2},2p_{3/2},0d_{5/2},3d_{5/2},1f_{7/2},0p_{1/2}\right)$,\\ $\left(2s_{1/2},3p_{3/2},0p_{1/2}\right)\leftrightarrow\left(3s_{1/2},1p_{1/2},0f_{7/2},0d_{3/2}\right)$,\\ $\left(3s_{1/2},1p_{1/2},0d_{3/2}\right)\leftrightarrow\left(2p_{1/2},0f_{5/2},1d_{3/2}\right)$,\\ $\left(2p_{1/2},1d_{3/2},0f_{5/2}\right)\leftrightarrow\left(3p_{1/2},1f_{5/2},2d_{3/2}\right)$ and\\ $\left(3p_{1/2},2d_{3/2},1f_{5/2}\right)\leftrightarrow\left(2f_{5/2},3d_{3/2}\right)$.

In table \ref{tab1}, in the absence of tensor interaction and when $C_{ps}$ becomes more negative, the system is becoming strongly attractive with the p-spin doublets $\left(1s_{1/2},2d_{5/2}\right)$ and $\left(1p_{3/2},0f_{5/2}\right)$  are degenerate sets for all values of $C_{ps}$. In the presence of spin symmetry, the increasing of $C_s$ value forces the system to become more repulsive and the considered states $1s_{1/2}$, $2p_{3/2}$, $0p_{1/2}$, $2f_{7/2}$ and $1f_{5/2}$ are not degenerate for all values of $C_s$ as shown in table \ref{tab2}.

\begin{table}[!t]
\caption{ The p-spin symmetric energy eigenvalues of the Yukawa potential for various values of $n$ and $\protect\kappa$ with $%
M=0.5fm^{-1}$, $V_0=1.0fm^{-1}$ and $a=0.1 fm^{-1}$ for various $C_{ps}$ values.}
\label{tab1}\vspace*{15pt} {\scriptsize 
\begin{tabular}{|c|c|c|c|c|c|}
\hline
\multicolumn{1}{|c|}{} & \multicolumn{5}{|c|}{$E_{n\kappa}(fm^{-1})$} \\%
[1.5ex] \hline
{} & {} & {} & {} & {} & {} \\[-1.0ex] 
$C_{ps}$ & $1s_{\frac{1}{2}}$ & $1p_{\frac{3}{2}}$ & $2d_{\frac{5}{2}}$ & $%
0f_{\frac{5}{2}}$ & $1g_{\frac{7}{2}}$ \\[2.5ex] \hline
-40 & -39.49974424 & -39.500000000 & -39.49974424 & -39.500000000 & 
-39.48746399 \\[1ex] 
& -8.220255757 & - & -8.220255757 & - & -0.611013166 \\[1ex] \hline
-35 & -34.49970674 & -34.500000000 & -34.49970674 & -34.500000000 & 
-34.48562441 \\[1ex] 
& -7.220293258 & - & -7.220293258 & - & -0.587472039 \\[1ex] \hline
-30 & -29.49965635 & -29.500000000 & -29.49965635 & -29.500000000 & 
-29.48315171 \\[1ex] 
& -6.220343648 & - & -6.220343648 & - & -0.564564029 \\[1ex] \hline
-25 & -24.49958505 & -24.500000000 & -24.49958505 & -24.500000000 & 
-24.47965078 \\[1ex] 
& -5.220414947 & - & -5.220414947 & - & -0.542684245 \\[1ex] \hline
-20 & -19.49947642 & -19.500000000 & -19.49947642 & -19.500000000 & 
-19.47431082 \\[1ex] 
& -4.220523578 & - & -4.220523578 & - & -0.522643493 \\[1ex] \hline
-15 & -14.49929074 & -14.500000000 & -14.49929074 & -14.500000000 & 
-14.46516171 \\[1ex] 
& -3.220709264 & - & -3.220709264 & - & -0.506411895 \\[1ex] \hline
-10 & -9.498900933 & -9.500000000 & -9.498900933 & -9.500000000 & 
-9.445829737 \\[1ex] 
& -2.221099067 & - & -2.221099067 & - & -0.500363156 \\[1ex] \hline
-5 & -4.497559159 & -4.500000000 & -4.497559159 & -4.500000000 & -4.376764780
\\[1ex] 
& -1.222440841 & - & -1.222440841 & - & -0.544047403 \\[1ex] \hline\hline
\end{tabular}%
} \vspace*{-1pt}
\end{table}

\begin{table}[!t]
\caption{ The spin symmetric energy eigenvalues of the Yukawa potential for various values of $n$ and $\protect\kappa$ with $%
M=0.5fm^{-1}$, $V_0=1.0fm^{-1}$ and $a=0.1 fm^{-1}$ for various $C_{s}$ values.}
\label{tab2}\vspace*{15pt} {\scriptsize 
\begin{tabular}{|c|c|c|c|c|c|}
\hline
\multicolumn{1}{|c|}{} & \multicolumn{5}{|c|}{$E_{n\kappa}(fm^{-1})$} \\%
[1.5ex] \hline
{} & {} & {} & {} & {} & {} \\[-1.0ex] 
$C_{s}$ & $1s_{\frac{1}{2}}$ & $1p_{\frac{3}{2}}$ & $0p_{\frac{1}{2}}$ & $%
2f_{\frac{7}{2}}$ & $1f_{\frac{5}{2}}$ \\[2.5ex] \hline
5 & 4.489714770 & 4.497433787 & 4.476781001 & 4.500000000 & 4.405380777 \\%
[1ex] 
& 0.839696995 & 1.382566213 & 0.728624405 & - & 0.721515775 \\[1ex] \hline
10 & 9.495503204 & 9.498876227 & 9.489875804 & 9.500000000 & 9.459363733 \\%
[1ex] 
& 1.128026208 & 2.381123773 & 0.850664737 & - & 0.702015577 \\[1ex] \hline
15 & 14.497121670 & 14.499280530 & 14.493522080 & 14.500000000 & 14.47405194
\\[1ex] 
& 1.420525390 & 3.380719471 & 0.982153599 & - & 0.721810125 \\[1ex] \hline
20 & 19.497883350 & 19.499470880 & 19.495236860 & 19.500000000 & 19.480933010
\\[1ex] 
& 1.713881360 & 4.380529119 & 1.115573949 & - & 0.749411816 \\[1ex] \hline
25 & 24.498326240 & 24.499581580 & 24.500000000 & 24.500000000 & 24.48492767
\\[1ex] 
& 2.007556118 & 5.380418419 & 1.249712246 & - & 0.779899919 \\[1ex] \hline
30 & 29.498615850 & 29.499653970 & 29.496885470 & 29.500000000 & 29.487537840
\\[1ex] 
& 2.301384153 & 6.380346026 & 1.384195613 & - & 0.811772503 \\[1ex] \hline
35 & 34.498820020 & 34.499705010 & 34.497344920 & 34.500000000 & 34.489377180
\\[1ex] 
& 2.595297632 & 7.380294988 & 1.518871297 & - & 0.844415924 \\[1ex] \hline
40 & 39.498971690 & 39.499742930 & 39.497686230 & 39.500000000 & 39.490743280
\\[1ex] 
& 2.889263601 & 8.380257072 & 1.653665117 & - & 0.877532579 \\%
[1ex] \hline\hline
\end{tabular}%
} \vspace*{-1pt}
\end{table}
\section{Concluding Remarks}
\label{C6}  
In this paper, we have obtained the approximate bound states of a Dirac particle confined to the field of the Yukawa potential and the tensor Coulomb-type interaction in the form of  $-A/r$. We used the asymptotic iteration method to obtain the energy eigenvalues and wave functions in closed form in the presence of the spin and the p-spin symmetries. In the presence of p-spin symmetry, some numerical values of the energy levels are calculated in table 1 for various values of $C_{ps}=-5 fm^{-1}$ to $C_{ps}=-40 fm^{-1}$. Also, in the presence of spin symmetry, the numerical energy levels are calculated in table 2 for various values of $C_{s}=5 fm^{-1}$ to $C_{s}=40 fm^{-1}$. Obviously, the degeneracy between the members of spin doublets and p-spin doublets is removed by the tensor interaction. The spin and p-spin spectra of the present potential is identical to those ones obtained in the previous works $\cite{NEW1, BJ33, BJ34, BJ35}$. We should remark that the present approximation is valid only for the lowest orbital states $\cite{BJ39, BJ40, BJ41}$. Finally, the relativistic spin symmetry in the absence of tensor interaction $-A/r$ and when $C_{s}=0 fm^{-1}$can be reduced to the non-relativistic solution for the Yukawa potential.

\begin{flushleft}
\textbf{\LARGE Acknowledgements}\\
\end{flushleft}
We wish to thank the kind referees for their useful suggestions and critics which have greatly improved this paper. S.M. Ikhdair thanks the partial support provided by the Scientific and Technological Research Council of Turkey (T\"{U}B$\dot{I}$TAK). B. J. Falaye acknowledges the efforts of Prof. Oyewumi,  K. J. for his encouragements.

\end{document}